\input{aipcheck}

\documentclass[
    ,final            % use final for the camera ready runs
  ]
  {aipproc}

\usepackage[]{graphicx}
\usepackage{amssymb}

\layoutstyle{8x11single}

\newcommand{\be}{\begin{equation}}
\newcommand{\ee}{\end{equation}}
\newcommand{\bea}{\begin{eqnarray}}
\newcommand{\eea}{\end{eqnarray}}

\begin{document}

\title{Higher-order brane gravity models}


\author{Mariusz P. D\c{a}browski and Adam Balcerzak}{
  address={Institute of Physics, University of Szczecin, Wielkopolska 15, 70-451 Szczecin, Poland}
%  email={mpdabfz@wmf.univ.szczecin.pl}
}

%\author{<author2>}{
%  address={<common address for author2 and author3>}
%}

%\author{<author3>}{
%  address={<common address for author2 and author3>}
%  ,altaddress={<author1 address>} % additional visiting address
%}

\begin{abstract}
 We discuss a very general theory of gravity, of which Lagrangian is an arbitrary function 
 of the curvature invariants, on the brane. In general, the formulation of
the junction conditions (except for Euler characteristics such as Gauss-Bonnet term) leads to
the powers of the delta function and requires regularization. We suggest the way to avoid such
a problem by imposing the metric and its first derivative to be regular at the brane, the second
derivative to have a kink, the third derivative of the metric to have a step function
discontinuity, and no sooner as the fourth derivative of the metric to give the delta function
contribution to the field equations. Alternatively, we discuss the reduction of the fourth-order
gravity to the second order theory by introducing extra scalar and tensor fields: the scalaron
and the tensoron. In order to obtain junction conditions we apply two methods: the application of
the Gauss-Codazzi formalism and the application of the generalized Gibbons-Hawking boundary terms
which are appended to the appropriate actions. In the most general case we derive junction conditions
without assuming the continuity of the scalaron and the tensoron on the brane. The derived junction
conditions can serve studying the cosmological implications of the higher-order brane gravity models.  
\end{abstract}

\maketitle

%%%%%%%%%%%%%%%%%%%%%%%%%%%%%%%%%%%%%%%%%%%%
%% MAINMATTER
%%%%%%%%%%%%%%%%%%%%%%%%%%%%%%%%%%%%%%%%%%%%
\section{1. Introduction}

In recent years there have been growing interest both in brane universes \cite{RS,brane1}
 and in the higher-order gravity theories of which the simplest is $f(R)$ gravity \cite{f(R)}. In this talk we are going to combine both ideas and formulate the higher-order gravities on the brane. It emerges that the formulation is a bit non-trivial, since one faces ambiguities of the quadratic delta function contributions to the field equations. We will say how to avoid these problems and show how the Israel junction conditions for such higher-order brane gravity models can be formulated.

\section{2. Fourth-order gravities.}

When one considers the general gravity theories (e.g. \cite{clifton,braneR2}):
\be
\label{XYZ}
S = \chi^{-1} \int d^D \sqrt{-g} f(X,Y,Z)
\ee
in a D-dimensional spacetime ($\chi =$ const.), where $X,Y,Z$ are curvature invariants
\be
X = R,\hspace{0.4cm} Y= R_{ab}R^{ab},\hspace{0.4cm} Z= R_{abcd}R^{abcd},
\ee
then one immediately faces the 4th order field equations, except
when they reduce to the theories with Euler densities of the n-th order $I^{(n)}$
\cite{lovelock}
\bea
\label{euler}
S = \int_M d^D x \sqrt{-g} \sum_n \kappa_n I^{(n)}~,
\eea
the lowest of them being the cosmological constant $I^{(0)} = 1$ ($\kappa_0 = -2\Lambda(2\kappa^2)^{-1} = -2\Lambda/16 \pi G$), the Ricci scalar $I^{(1)} =R$ ($\kappa_1 = (2\kappa^2)^{-1}$), and the Gauss-Bonnet density $I^{(2)} = R{GB} = R^2 - 4 R_{ab}R^{ab} + R_{abcd}R^{abcd}$ ($\kappa_2=\alpha(2\kappa^2)^{-1}$, $\alpha=$ const.).

However, the theories based on the Lagrangians which are the functions of the Euler densities
such as
\bea
f(R) = f(X), \hspace{1.5cm} f(R_{GB}) = f(Z-4Y+X^2)~,\hspace{1.5cm} f = f(I^{(n)})
\eea
are again fourth-order.

\section{3. Formulation of the 4th order gravities on the brane - Israel formalism.}

In the context of the recent interest in string/M-theory, it is interesting to formulate the general gravity theories (\ref{XYZ}) within the framework of the brane models \cite{PRD08}. The full brane action for such a theory reads as
\bea
\label{XYZB}
S &=& \chi^{-1} \int_{M} d^{D}x \sqrt{-g} f(X,Y,Z) + S_{brane} + S_{m}~,
\eea
with the total energy-momentum tensor
\begin{eqnarray}
\label{Tab}
T_a^{~b}=T_{a}^{~b~-}\theta(-w) + T_{a}^{~b~+}\theta(w) +
\delta(w)S_a^{~b},
\end{eqnarray}
where $S_a^{~b}$ is the energy-momentum tensor on the brane, and
$T_{a}^{~b~\pm}$ are the energy-momentum tensors on the both sides of the brane, 
$\theta(w)$ is the Heaviside step function, and $\delta(w)$ is the Dirac delta function.

We assume Gaussian normal coordinates, i.e.,
$(\mu,\nu = 0, 1, 2,\ldots,D-2;w=D)$
\begin{eqnarray}
\label{bm}
ds^2=g_{ab} dx^a dx^b = \epsilon dw^2+h_{\mu\nu}dx^{\mu}dx^{\nu}~,
\end{eqnarray}
where $\epsilon = \vec{n} \cdot \vec{n} = +1$ for a spacelike hypersurface,
$\epsilon= -1$ for a timelike hypersurface, and $h_{ab} = g_{ab} - \epsilon n_a n_b$
is a projection tensor onto a $(D-1)$-dimensional hypersurface, $\vec{n}$ is the normal vector to the hypersurface. In these coordinates the extrinsic
curvature is
\begin{eqnarray}
K_{\mu\nu}=-{1\over 2}{\partial h_{\mu\nu}\over\partial w}~,
\end{eqnarray}
and the Gauss-Codazzi equations read \cite{brane2}
\begin{eqnarray}
\label{GC}
R_{w\mu w\nu}&=& {\partial K_{\mu\nu}\over \partial w}+K_{\rho\nu}K^{\rho}_{\,\,\mu},  \\
R_{w\mu\nu\rho}&=&\nabla_{\nu}K_{\mu\rho}-\nabla_{\rho}K_{\mu\nu}, \\
R_{\lambda\mu\nu\rho}&=&~^{(D-1)} R_{\lambda\mu\nu\rho}+
\epsilon\left[K_{\mu\nu}K_{\lambda\rho}
-K_{\mu\rho}K_{\lambda\nu}\right]~.
\end{eqnarray}
In the standard Israel approach \cite{israel66} one assumes that at the brane position $w=0$:
\begin{eqnarray}
\label{cont1}
h^{-}_{\mu\nu} &=& h^{+}_{\mu\nu}~,\\
\label{cont2}
h^{-}_{\mu\nu,w} & \neq & h^{+}_{\mu\nu,w}~, \hspace{0.5cm} K^{-}_{\mu\nu}
\neq K^{+}_{\mu\nu}~,
\end{eqnarray}
i.e., the {\it metric is continuous} but it has a kink, its first derivative has {\it a step function} discontinuity, and its second derivative gives the {\it delta function} contribution.

In terms of $\theta(w)$ and $\delta(w)$ functions this is equivalent to
\begin{eqnarray}
h_\mu{_\nu}(w) &=& h^{-}_{\mu\nu}(w) \theta(-w) + h^{+}_{\mu\nu}(w)
\theta(w) ~,\\
{\partial h_{\mu\nu}} \over \partial w &=&  {\partial h^{+}_{\mu\nu} \over
\partial w} \theta(-w) +  {\partial h^{-}_{\mu\nu} \over \partial w} \theta(w)~, \\
{\partial{^2} {h_\mu{_\nu}} \over \partial w{^2}}&=&  {\partial{^2} h^{-}_{\mu\nu}
\over \partial w{^2}} \theta(-w) +  {\partial{^2} h^{+}_{\mu\nu} \over \partial w{^2}}
\theta(w) \nonumber \\ &+& \left( {\partial h^{-}_{\mu\nu} \over \partial w} -
{\partial h^{+}_{\mu\nu} \over \partial w} \right)\delta(w)~.
\end{eqnarray}
For the standard brane models with the Einstein-Hilbert action in the bulk
\bea
S = \frac{1}{2\kappa^2}\int_{M} d^{D}x \sqrt{-g} R + S_{brane} +
S_{m}
\eea
the field equations read as \cite{brane2}
\begin{eqnarray}
\label{Gww}
G^w_{~w}&=&-{1\over 2}~^{(D-1)}R+{1\over 2}\epsilon\left[K^2-Tr(K^2)\right]=\kappa^2 T^w_{~w}, \\
\label{Gwm}
G^w_{~\mu}&=&\epsilon\left[\nabla_{\mu}K-\nabla_{\nu}K^{\nu}_{\,\,\mu}\right]=\kappa^2
T^w_{~\mu}, \\
\label{Gmm}
G^{\mu}_{~\nu}&=&~^{(D-1)}G^{\mu}_{~\nu}
             +\epsilon\left[{\partial K^{\mu}_{~\nu}\over\partial w}-\delta^{\mu}_{~\nu}
{\partial K\over\partial w}\right]\\
&+& \epsilon\left[-
K K^{\mu}_{~\nu}+{1\over 2}\delta^{\mu}_{~\nu}Tr(K^2)+{1\over 2}\delta^{\mu}_{~\nu}
K^2\right]=\kappa^2 T^{\mu}_{~\nu}~.\nonumber
\end{eqnarray}
and in the limit $ \lim_{w \to 0} \int_{-w}^{w}$, which ``fishes out'' the delta function contributions, one gets the {\it standard Israel junction conditions} as \cite{brane2}:
\begin{eqnarray}
\label{jcE}
\epsilon \{ [K^{\mu}_{~\nu}]-\delta^{\mu}_{~\nu}[K]\} &=& \kappa^2 {S}^{\mu}_{~\nu},
\hspace{0.5cm} [K^{\mu}_{~\nu}] \equiv K^{\mu~+}_{~\nu}-K^{\mu~-}_{~\nu}.
\end{eqnarray}
By $[X] = X^+ + X^-$ we define a jump of an appropriate quantity $X$ at the brane. 

However, for the general $f(X,Y,Z)$ theory on the brane, the standard continuity relations (\ref{cont1})-(\ref{cont2}) do not work. This can be seen from the field equations
of the action (\ref{XYZ})
\begin{eqnarray}
\label{XYZ1}
P_{a b}&=&\frac{\chi}{2} T_{a b}, \\
\label{XYZ2}
P^{a b} &=& -\frac{1}{2} f g^{a b} + f_X R^{a b}+2 f_Y R^{c (a} {R^{b)}}_{c}+2
f_Z R^{e d c (a} {R^{b)}}_{c d e} \nonumber \\ &+& f_{X; c d}(g^{a
b} g^{c d}-g^{a c} g^{b d}) + \square (f_Y R^{a b}) + g^{a b} (f_Y
R^{c d})_{;c d} \nonumber \\ &-& 2 (f_Y R^{c (a})_{;\; \; c}^{\;
b)}-4 (f_Z R^{d (a b) c})_{;c d},
\end{eqnarray}
where $f_X = {\partial f / \partial X}$ etc.

Take, for example, the square of the Ricci scalar
\begin{eqnarray}
R&=&~^{(D-1)}R+\epsilon\left[2h^{\mu\nu}{\partial K_{\mu\nu}\over\partial w}
+3Tr(K^2)-K^2\right]~,\nonumber
\end{eqnarray}
where $K\equiv K^{\mu}_{\,\,\mu}$ , $Tr(K^2)\equiv K^{\mu\nu}K_{\mu\nu}$,
and appropriately, of the Ricci tensor, and of the Riemann tensor. These squares produce the terms of the type
\bea
\label{terms}
{\partial^2 h^{\mu \nu} \over \partial^2 w}{\partial K_{\mu \nu} \over \partial w},
   {\partial K_{\mu \nu} \over \partial w}{\partial K^{\mu \nu} \over \partial w},
   \left({\partial K \over \partial w}\right)^2~,
\eea
which are proportional to $\delta^2(w)$, and so they are {\it ambiguous}.

Amazingly, all these ambiguous terms cancel each other exactly in the case of the Euler densities \cite{meissner01}. In fact, the junction conditions for one of the Euler densities -- the Gauss-Bonnet density, were already obtained as \cite{deruelle00,davis}
\bea
2 \alpha \left( 3 [J_{\mu\nu}] - [J] h_{\mu\nu}
- 2 [P]_{\mu\rho\nu\sigma} [K]^{\rho\sigma} \right)
+ [K_{\mu\nu}] - [K] h_{\mu\nu} = - \kappa^2 S_{\mu\nu}~,
\eea
where
\bea
P_{\mu\rho\nu\sigma} &=& R_{\mu\rho\nu\sigma} + 2 h_{\mu[\sigma}R_{\nu]\rho}
+ 2 h_{\rho[\nu}R_{\sigma]\mu}
+ R h_{\mu[\nu}h_{\sigma]\rho}~, \\
J_{\mu\nu} &=& \frac{1}{3} \left( 2KK_{\mu\sigma}K^{\sigma}_{\nu} +
K_{\sigma\rho}K^{\sigma\rho}K_{\mu\nu} -
2K_{\mu\rho}K^{\rho\sigma}K_{\sigma\nu}
- K^2 K_{\mu\nu} \right)~.
\eea
In the limit $\alpha \to 0$, they just give Einstein-Hilbert action junction conditions (\ref{jcE}).

In view of the ambiguities of the terms in (\ref{terms}), we find two ways to formulate the junction conditions for general $f(X,Y,Z)$ theories on the brane.

\subsection{A. Smoothing out the continuity conditions for the metric tensor at the brane}

In order to do that we impose more regularity onto the metric tensor at the brane position, i.e., we consider {\it a singular hypersurface of the order three} \cite{israel66} which fulfills the conditions (compare (\ref{cont1})-(\ref{cont2}))
\begin{eqnarray}
\label{hh1}
h^{-}_{\mu\nu} &=& h^{+}_{\mu\nu}~,\\
\label{hh2}
h^{-}_{\mu\nu,w} & = & h^{+}_{\mu\nu,w}~, \hspace{0.5cm} K^{-}_{\mu\nu}
= K^{+}_{\mu\nu}~,\\
\label{hh3}
h^{-}_{\mu\nu,ww} &=& h^{-}_{\mu\nu,ww}~, \hspace{0.5cm} K^{-}_{\mu\nu,w} =
K^{+}_{\mu\nu,w}~,\\
\label{hh4}
h^{-}_{\mu\nu,www} & \neq & h^{+}_{\mu\nu,www}~, \hspace{0.5cm} K^{-}_{\mu\nu,ww}
\neq K^{+}_{\mu\nu,ww}~,
\end{eqnarray}
i.e., the metric and its first derivative are regular, the {\it second derivative of the
metric is continuous}, but possesses a kink, the third derivative of the metric
has {\it a step function} discontinuity, and no sooner than the fourth derivative of the
metric on the brane produces the {\it delta function} contribution.

The physical interpretation as put in terms of the second-order theory can be that there is a jump of the first derivative of the energy-momentum tensor (e.g. jump of a pressure gradient) at the brane. 

In his seminal work, Israel \cite{israel66} proposed {\it a singular hypersurface of order two}, which physically corresponded to a boundary surface characterized by a jump of the energy-momentum tensor (e.g. a boundary surface separating a star from the surrounding vacuum) which was characterized by
\begin{eqnarray}
\label{hhb1}
h^{-}_{\mu\nu} &=& h^{+}_{\mu\nu}~,\\
\label{hhb2}
h^{-}_{\mu\nu,w} & = & h^{+}_{\mu\nu,w}~, \hspace{0.5cm} K^{-}_{\mu\nu}
= K^{+}_{\mu\nu}~,\\
\label{hhb3}
h^{-}_{\mu\nu,ww} &\neq& h^{-}_{\mu\nu,ww}~, \hspace{0.5cm} K^{-}_{\mu\nu,w} = K^{+}_{\mu\nu,w}~,
\end{eqnarray}
i.e., the metric is regular, the {\it first derivative of the
metric is continuous}, but possesses a kink, the second derivative of the metric has {\it a step function} discontinuity, and the third derivative of the metric on the brane produces the {\it delta function} contribution.

\begin{figure}[h]
%\label{SFS}
%\caption{\scriptsize BB to SFS evolution}
%\begin{center}
\includegraphics[width=8cm]{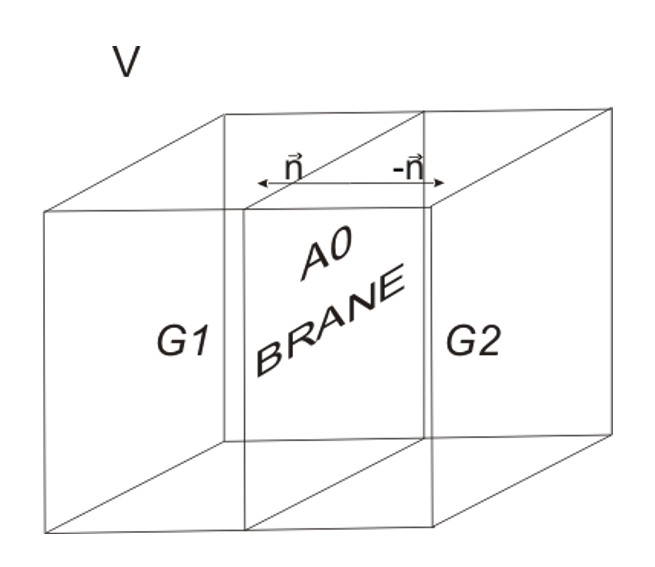}
\caption{A schematic picture illustrating the domains of integration
        used in derivation of the junction conditions. Here $V= G1+G2$ is the 
        total volume, $G1$, $G2$ - are the
left-hand-side and the right-hand-side bulk volumes which are
separated by the brane, $A1=
\partial G1 + A0$, $A2= \partial G2 - A0$ are the boundaries of
these volumes, and $A0$ is the brane which orientation is given by
the direction of the normal vector $\vec{n}$.}
        \label{fig1}
%\end{center}
\end{figure}

The appropriate junction conditions can be obtained as follows.
We rewrite the field equations (\ref{XYZ1})-(\ref{XYZ2}) as
\begin{eqnarray}
\label{Wabd}
\sqrt{-g}C_{ab}{W^{abd}}_{;d} + \sqrt{-g}C_{ab}V^{ab} =
{\chi \over 2} T^{ab}C_{ab}\sqrt{-g}~,
\end{eqnarray}
where we have introduced is an arbitrary tensor field $C_{ab}$, and
\begin{eqnarray}
    W^{abd}&=&f_{X; c }(g^{a b} g^{c d}-g^{(a c} g^{b) d}) + (f_Y
    R^{ab})^{;d} \\ \nonumber
    &+& g^{ab}(f_Y R^{cd})_{;c}  -2 (f_Y R^{d(a})^{;b)}
    - 4(f_Z R^{d(ab)c})_{;c}~,\\
    V^{ab} &=& -\frac{1}{2} f g^{a b} + f_X R^{a b}+2 f_Y R^{c (a}
    {R^{b)}}_{c} \nonumber \\
    &+& 2 f_Z R^{e d c (a} {R^{b)}}_{c d e}~,
\end{eqnarray}
contain third derivatives of the metric giving a step function discontinuity, so that ${W^{abd}}_{;d}$ is proportional to $\delta(w)$. Then, we integrate both sides of the formula (\ref{Wabd}) over the volume $V$ which contains the
following parts (cf. Fig. \ref{fig1}): $G1$, $G2$ - are the
left-hand-side and the right-hand-side bulk volumes which are
separated by the brane, $A1=
\partial G1 + A0$, $A2= \partial G2 - A0$ are the boundaries of
these volumes, and $A0$ is the brane which orientation is given by
the direction of the normal vector $\vec{n}$. 

We have
\begin{eqnarray}
\int_{G1+G2}{\sqrt{-g}C_{ab}{W^{abd}}_{;d} d\Omega}  
+ \int_{G1+G2} {\sqrt{-g}C_{ab}V^{ab}d\Omega}
=\int_{G1+G2}{{\chi \over 2}T^{ab}C_{ab}\sqrt{-g}d\Omega}~, 
\end{eqnarray}
and so
\begin{eqnarray}
&& \int_{G1+G2}\sqrt{-g}(C_{ab}W^{abd})_{;d} d\Omega
- \int_{G1+G2}\sqrt{-g}C_{ab;d}W^{abd}
d\Omega  
+ \int_{G1+G2}\sqrt{-g}C_{ab}V^{ab}d\Omega  \nonumber \\ 
&=& \int_{G1}{\chi \over 2} T^{ab}C_{ab}\sqrt{-g}d\Omega + \int_{G2}{\chi \over 2}
T^{ab}C_{ab}\sqrt{-g}d\Omega 
+ \int_{A0}{\chi \over 2} S^{ab}C_{ab}\sqrt{-\gamma}d\sigma~,
\end{eqnarray}
of which the first term can be integrated out to a boundary A1+A2 and then the limit $V \to A0$ (or $ \lim_{w \to 0} \int_{-w}^{w}$ in Gaussian coordinates) is taken.

The final form of the junction conditions which generalize (\ref{jcE}) onto the fourth-order gravity are
\begin{eqnarray}
\label{ws}
    [W]^{abd}n_d - {\chi \over 2} S^{ab} &=& 0~, \hspace{.3cm} [W]^{abd} = W^{abd+} - W^{abd-}.
\end{eqnarray}

It is remarkable that these junction conditions involve the higher derivatives of the scale factor. To see this take for example $f(X,Y,Z)=f(R)$ theory in $D=5$ dimensions with metric
\begin{eqnarray}
\label{bw1}
ds^2=-dt^2
+ a^2(t,w)[dr^2 +r^2(d\Theta^2 +sin^2\Theta d\phi^2)]+dw^2~~.
\nonumber
\end{eqnarray}
The junction conditions (\ref{ws}) give a jump of the third derivative of $a(t,w)$, as expected
\begin{eqnarray}
    [a'''] &=&  {\chi \over 2}{a_0} {p_0}~, \\
    p_0&=& \rho_0~,
\end{eqnarray}
where $(\ldots)' = \partial / \partial w$, $a_0=a(w=0)$, and the
brane energy-momentum tensor is $S_{\mu}^{\nu} =
(-\rho_0,p_0,p_0,p_0)$.

\subsection{B. Reduction to an equivalent 2nd order theory}

Yet another way to obtain the junction conditions is the reduction of the action (\ref{XYZ}) to a second-order action. This gives equivalent junction conditions, though at the expense of introducing a new tensor field $H^{abcd}$ (tensoron). In fact, starting from the action \cite{kijowski}
\bea \label{r} S_{G} &=& \chi^{-1} \int_{M} d^{D}x \sqrt{-g}
f(g_{ab},R_{abcd}).
\eea
we may transform to an equivalent 2nd order action in the form
\bea \label{equiv}
 S_{I} = \chi^{-1} \int_{M} d^{D}x \sqrt{-g} \{ H^{ghij}(R_{ghij}-
 \phi_{ghij}) + f(g_{ab},\phi_{cdef}) \}~,
 \eea
where
\bea \label{H} H^{ghij} \equiv {\partial f(g_{ab},\phi_{abcd}) \over
\partial \phi_{ghij}}~, \hspace{1.cm} det \left[{\partial^2 f(g_{ab},\phi_{abcd}) \over
\partial \phi_{ghij} \partial \phi_{klmn}} \right] \neq 0.
\eea
This transition for $f(R)$ theory requires a new scalar $H=f'(Q)$ (a scalaron) with the condition that $f''(Q) \neq 0$, and the equation of motion $Q=R$. Similarly, for $f(R_{GB})$ theory, one defines a scalar $H=f'(A)$, with the equation of motion  $A=R_{GB}$).
In order to get junction conditions, we have to slightly redefine the tensoron
\bea
A^{abcd}={1 \over 2} \{H^{acdb} &+& H^{abdc}-H^{cbda} - H^{acbd} - H^{abcd}+H^{cbad}\}
\eea
which in a particular case of $f(X,Y,Z)$ theory takes the form
\bea
A^{abcd}= f_{X}(g^{ad} g^{cb}-g^{cd} g^{ba}) + f_{Y}(2R^{ad}g^{bc} - R^{cd}g^{ba} -R^{ba}g^{cd}) + 4f_{Z}R^{acbd}~.
\eea

The field equations for an equivalent action (\ref{r}) read as
\bea \label{em1}
R_{ghij} &=& - {\partial V(g_{ab},H^{cdef}) \over \partial H^{ghij}}~, \\
\label{em2} {1\over 2} g^{ab}f &+& {\partial f \over \partial
g_{ab}} + H^{becd} \phi ^{a}_{~ecd}(g_{ab},H^{klmn})
+\{A^{(ab)cd}\}_{;dc}
 = - {\chi \over 2} T^{ab}~,
\eea
where
\bea
V(g_{ab},H^{cdef}) = - H^{hgij} \phi_{ghij}(g_{ab},H^{cdef}) +
f(g_{ab},\phi_{klmn}(g_{ab},H^{cdef})) 
\eea
In fact, the possibility to express the fields $\phi_{abcd}$ as a function of
$g_{ab}$ and $H^{cdef}$ is guaranteed by the condition
(\ref{H}) (which is an analogue of the condition $f''(Q) \neq 0$).

One can show that junction conditions of the second-order theory are equivalent to junction 
conditions of the fourth-order theory \cite{PRD08}.

 Applying the same method as in the previous case (i.e. taking the limit of $V \to A0$) we notice that the first three terms of (\ref{em2}) do not give any contribution to the junction conditions (since they do not contain delta functions at all) which now have the form:
\bea
\label{jc1}
[{A^{(ab)cd}}_{;d}]n_{c} = - {\chi \over 2} S^{ab}~.
\eea
Assuming that
\bea \label{f}
f(g_{ab},\phi_{abcd})=f({\phi_{ab}}^{ab},{\phi_{acb}}^{c}{\phi^{acb}}_{c},\phi_{abcd}
\phi^{abcd}),
\eea
we can get the same result as in the 4th theory
\bea \label{equivH}
{[A^{(ab)cd}}_{;d}]n_{c}&=&{[A^{(ab)cd}}_{;c}]n_{d} = [- \{f_{X; c
}(g^{ab} g^{c d}-g^{c(a} g^{b)d}) \nonumber \\  &+& (f_Y  R^{ab})^{;d} + g^{ab}(f_Y R^{cd})_{;c} \\
\nonumber -2(f_YR^{d(a})^{;b)} &-& 4(f_Z R^{d(ab)c})_{;c}\}]n_{d}=
-[W^{abd}]n_{d}. \eea

Similar approach was used for less-general $f(R)$ theories of gravity on the brane by 
Borzeszkowski and Frolov \cite{borzeszkowski}; Parry at al. \cite{branef(R)}, Deruelle et al. \cite{deruelle07}, and for $f(X,Y,Z) = aX^2 + bY + cZ$ ($a,b,c =$ const.) theories by Nojiri and Odintsov \cite{braneR2}.

\section{4. Formulation of the 4th order gravities on the brane - Gibbons-Hawking Boundary Terms}

In this approach, following the idea of Gibbons and Hawking \cite{GH}, we do not assume any vanishing of the first derivative of the variation of the metric tensor $\delta g_{ab;c}$ on the boundary of the integration volume while using the variational principle. Strictly speaking, only the assumption of the vanishing of the normal derivative of the variation of the metric tensor $\delta g_{ab,w}$ is required. Instead, we postulate that some extra terms to the action are added and that these terms ``kill'' the first derivatives of the metric variation. These terms are called Gibbons-Hawking boundary terms now. In fact, the Gibbons-Hawking boundary term for the Einstein-Hilbert action is composed of the trace of the extrinsic curvature and it was found by Gibbons and Hawking themselves \cite{GH}. Then, for the action being the combination of the square of the Weyl tensor and an arbitrary function of the scalar curvature they were found by Hawking and Lutrell \cite{lutrell} and Barrow and Madsen \cite{madsen}.
For the Gauss-Bonnet and other Lovelock densities they were found by Bunch \cite{bunch81}, Mueller-Hoissen and Myers \cite{surface}, Davis \cite{davis} and Gravanis and Willinson \cite{gravanis}. The boundary terms for the action being an arbitrary function of the curvature invariants were found by Barvinsky and Solodukhin \cite{barvinsky}.

For the theories which are of interest for this talk, the Gibbons-Hawking boundary terms have
the following form \cite{JCAP09}: for the $f(R)$ theory the term reads as 
\begin{eqnarray}
\label{gib}
S_{GH,p}= -2(-1)^{p}\epsilon \int_{\partial M_p} \sqrt{-h} H K
d^{D-1}x~,
\end{eqnarray}
where $H=f'(Q)$ is the scalaron,
while for the $f(X,Y,Z)$ theory it reads as
\bea
\label{gib2}
S_{GH,p} =
 - (-1)^{p} \int_{\partial M_p}
d^{D-1}x \sqrt{-h} A^{(ab)cd}n_{c} n_{d}\mathcal{L}_{\vec{n}}g_{ab}~,
\eea
where $A^{(ab)cd}$ is the tensoron.

Using the method of the boundary terms we derived the most general Israel junction conditions for $f(R)$ theory as \cite{JCAP09}:
\begin{eqnarray}
\label{jc2}
[K]&=& 0~, \\
\label{jc21}
S^{ab}n_{a}n_{b}&=& 0~, \\
\label{jc22}
S^{ab}h_{ac}n_{b}&=& 0~, \\
\label{jc23}
-(D-1)[H_{;c}n^{c}]-D[H]K &=& \epsilon {\chi \over 2} S^{ab}h_{ab}~,\\
\label{jc24}
-h_{ab}[H_{;c}n^{c}]-[H]Kh_{ab} &+& [HK_{ab}] \\
&=&
\epsilon {\chi \over 2}S^{cd}h_{ca}h_{db}. \nonumber
\end{eqnarray}
A generality of these conditions refers to the fact that no assumption about the 
continuity of the scalaron on the brane has been made. They reduce to the conditions already obtained in the literature, if one assumes $[H] =0$ \cite{deruelle07}.

On the other hand, the most general Israel junction conditions for the $f(X,Y,Z)$ theory,
with no assumption about the continuity of the tensoron on the brane, are \cite{JCAP09}:
\bea
\label{JCXYZ1}
&&[KA^{(ab)cd}] n_{c} n_{d} + [\mathcal{L}_{\vec{n}}A^{(ab)cd}] n_{c} n_{d}
\\ \nonumber &-& \epsilon[A^{(ab)cd}K_{cd}] - g^{ab}[A^{(ef)cd} K_{ef}]n_{c} n_{d}  \\
\nonumber &+& 2 \epsilon [D_{s}A^{(ef)cd}n_{c} n_{d}]h^{s}_{e}h^{(a}_{f}n^{b)}
-  2\epsilon[{A^{(ab)cd}}_{;(c}]n_{d)} = {\chi \over 2} S^{ab}~,
 \\
\label{JCXYZ2}
&& 
 n_{b} n_{c}[\mathcal{L}_{\vec{n}}g_{ad}]-
 n_{a} n_{c}[\mathcal{L}_{\vec{n}}g_{db}]-n_{b} n_{d}[\mathcal{L}_{\vec{n}}g_{ac}]
 +n_{a} n_{d}[\mathcal{L}_{\vec{n}}g_{cb}]=0~.
\eea
They reduce to the conditions (\ref{jc1}), if one assumes continuity of the tensoron on the brane 
\begin{equation}
[A^{(ab)cd}] = 0~.
\end{equation}

\section{5. Fourth-order gravities and statefinders}

We claim the fact that general $f(R,R_{ab}R^{ab},R_{abcd}R^{abcd})$ theories are fourth-order may have some advantageous consequences onto their observational verification
by the application of statefinder diagnosis of the universe.

In fact, statefinders are the higher-order characteristics of the universe expansion which go
beyond the Hubble parameter $H$ and the deceleration parameter $q$:
\bea
\label{hubb}
H &=& \frac{\dot{a}}{a}~,\hspace{0.5cm} q  =  - \frac{1}{H^2} \frac{\ddot{a}}{a} = - \frac{\ddot{a}a}{\dot{a}^2}~.
\eea
They can generally be expressed as ($i \geq 2$)
\bea
\label{dergen}
x^{(i)} &=& (-1)^{i+1}\frac{1}{H^{i}} \frac{a^{(i)}}{a} = (-1)^{i+1}
\frac{a^{(i)} a^{i-1}}{\dot{a}^{i}}~,
\eea
and the lowest order of them are known as:
jerk, snap ("kerk"), crack ("lerk")
\bea
\label{jerk}
j &=& \frac{1}{H^3} \frac{\dot{\ddot{a}}}{a} =
\frac{\dot{\ddot{a}}a^2}{\dot{a}^3}~, \hspace{0.5cm} k = -\frac{1}{H^4} \frac{\ddot{\ddot{a}}}{a} = -\frac{\ddot{\ddot{a}}a^3}{\dot{a}^4}~, l = \frac{1}{H^5} \frac{a^{(5)}}{a} =
\frac{a^{(5)} a^4}{\dot{a}^5}~,
\eea
and pop ("merk"), "nerk", "oerk", "perk" etc. \cite{statefind}.

In the case of the 4th order gravities, statefinders may become  powerful tools to constrain such theories observationally, since they enter observational relations
in the higher orders of redshift $z$ (see \cite{statefR} for non-brane case diagnosis). 

Apparently, a blow-up of statefinders may also be linked to an
emergence of exotic singularities in the universe \cite{blowup}.

\section{6. Conclusions}

We conclude the following:

\begin{itemize}
\item The formulation of the fourth-order gravity theories on the brane is non-trivial because of the powers of {\it delta function ambiguities}.

\item Two methods were applied: \\
A. {\it Smoothing out} the continuity conditions for the metric tensor at the brane; \\  
B. Reduction to an {\it equivalent} 2nd order theory.
\\In both cases the Israel {\it junction conditions} have been obtained and they {\it are} also mutually {\it equivalent}.

\item The method of the {\it GH boundary terms} was also applied and the most general junction conditions (with no continuity of the scalaron and tensoron on the brane assumed) were obtained that way, too.

\item Higher-order brane gravities contain {\it higher-order derivatives} of the geometric quantities (in a Friedmann model it is just the scale factor) which may manifest themselves in the {\it higher-order characteristics of expansion} such as statefinders (jerk, snap, lerk/crack, merk/pop). 
    
\item A blow-up of statefinders may be linked to an {\it emergence of exotic singularities} in the universe.
\end{itemize}

%%%%%%%%%%%%%%%%%%%%%%%%%%%%%%%%%%%%%%%%%%%%
%% Sample figure:
%%
%% The option [height=...] scales the picture to the given height,
%% without it it would be printed at its nominal size
%%%%%%%%%%%%%%%%%%%%%%%%%%%%%%%%%%%%%%%%%%%%

%\begin{figure}
%  \includegraphics[height=.3\textheight]{golfer}
%  \caption{Picture to fixed height}
%\end{figure}

%\section{<A section>}

%%%%%%%%%%%%%%%%%%%%%%%%%%%%%%%%%%%%%%%%%%%%%%%%
%% BACKMATTER
%%%%%%%%%%%%%%%%%%%%%%%%%%%%%%%%%%%%%%%%%%%%%%%%

\begin{theacknowledgments}
We acknowledge the support of the Polish Ministry of Science
and Higher Education grant No N N202 1912 34 (years 2008-10).
\end{theacknowledgments}




%\bibliography{sample}

%  \typeout{******************************************}
%  \typeout{** Please run "bibtex \jobname" to optain}
%  \typeout{** the bibliography and then re-run LaTeX}
%  \typeout{** twice to fix the references!}
%  \typeout{******************************************}

\end{document}